\documentclass[useAMS,usenatbib]{mn2e}

\usepackage[pdftex]{graphicx}
\usepackage{multirow}
\usepackage{aas_macros}
\usepackage{mathptmx}

\newcommand{\be}{\begin{equation}}
\newcommand{\ee}{\end{equation}}
\newcommand{\bm}{\bmath}
\newcommand{\B}{\mathbf}

\newcommand{\Nside}{{N_\rmn{side}}}
\newcommand{\bhat}[1]{\hat{\bm{#1}}}
\newcommand{\LCDM}{$\Lambda$CDM}
\newcommand{\lmax}{{\ell_{\max}}}
\newcommand{\eqref}[1]{(\ref{#1})}

\newcommand{\halfW}{0.45\textwidth}
\newcommand{\fullW}{0.95\textwidth}

\title[CMB mirror-parity after \textit{Planck}]{A close examination of cosmic microwave background mirror-parity after \textit{Planck}}

\author[A.~Ben-David and E.~D.~Kovetz]{Assaf Ben-David$^1$\thanks{E-mail: bendavid@nbi.dk} and Ely D.~Kovetz$^2$\thanks{E-mail: elykovetz@gmail.com} \\
$^1$Niels Bohr International Academy and Discovery Center, The Niels Bohr Institute, Blegdamsvej 17, DK-2100 Copenhagen \O, Denmark \\
$^2$Theory Group, Department of Physics and Texas Cosmology Center, The University of Texas at Austin, TX 78712, USA}

\begin{document}

\pagerange{\pageref{firstpage}--\pageref{lastpage}} \pubyear{2014}

\maketitle

\label{firstpage}

\begin{abstract}
Previous claims of significant evidence for mirror-parity in the large-scale cosmic microwave background (CMB) data from the \textit{Wilkinson Microwave Anisotropy Probe} (\textit{WMAP}) experiment have been recently echoed in the first study of isotropy and statistics of CMB data from \textit{Planck}. We revisit these claims with a careful analysis of the latest data available. We construct statistical estimators in both harmonic and pixel space, test them on simulated data with and without mirror-parity symmetry, apply different Galactic masks, and study the dependence of the results on arbitrary choices of free parameters. We confirm that the data exhibit evidence for odd mirror-parity at a significance which reaches as high as $\sim99$ per cent C.L., under some circumstances. However, given the bias exhibited by the pixel-based statistic on a masked sky, its sensitivity to the total power and the dependence of both pixel and harmonic space statistics on the particular form of Galactic masking and other a-posteriori choices, we conclude that these results are not in significant tension with the predictions of the concordance cosmological model.
\end{abstract}


\section{Introduction}
\label{sec:Introduction}

The importance of large scale symmetries in the cosmic microwave background (CMB), if those are shown to exist, cannot be overstated. From a theoretical standpoint, these scales provide us with a possible observational window to the very early Universe, and to the physics at very high energies. Deviations from the expectations of the concordance $\Lambda$ cold dark matter (\LCDM) cosmological model on these scales, such as the breaking of statistical isotropy, could indicate a special location, axis or direction, which in turn can point to the existence of exotic pre-inflationary relics or other effects in the pre-inflationary Universe.

Mirror-parity is an intriguing example of a symmetry which breaks statistical isotropy. As it involves a symmetry plane, it can be naturally attributed to various early Universe models predicting large-scale modulation of the CMB (\citealt{Gordon:2005jk}; \citealt*{Ackerman:2007sf}; \citealt{Hou:2010xy,Schmidt:2013ty}). Another interesting class of models that could induce this symmetry involves a finite topology for the Universe \citep*{de-Oliveira-Costa:1996qy}. In such models, statistical isotropy on large scales may be broken by the finite fundamental domain \citep{Levin:2002nr,Riazuelo:2004qv}. 

When searching for a mirror-parity symmetry plane on real data, however, care must be taken to avoid spurious effects that may induce even mirror-parity due to experimental systematics (some of which involve the ecliptic plane) or foregrounds (which dominate the Galactic plane). Also, as always, inherent biases and a-posteriori choices in the definition of the statistical estimators may lead to unfounded conclusions regarding anisotropies in the data. 

In this paper, we revisit recent claims of borderline significant detection of mirror-parity in the CMB (\citealt{Land:2005ys}; \citealt*{Ben-David:2012eu}; \citealt{Finelli:2012ij,Planck-Collaboration:2013oq}), in data from both the \textit{Wilkinson Microwave Anisotropy Probe} \citep[\textit{WMAP}; see e.g.][]{Bennett:2011lr} and \textit{Planck} \citep{Planck-Collaboration:2013kl} experiments. The different reports have been consistent so far in identifying two prominent directions in the sky, one which maximizes even mirror-parity and another corresponding to odd mirror-parity. The even mirror-parity direction found \citep{Land:2005ys,Ben-David:2012eu,Finelli:2012ij} coincides well with the direction of the CMB dipole \citep{Kogut:1993hl,Bennett:2003qd,Planck-Collaboration:2013rc}, and its reported significance was rather mild. Meanwhile, the direction in which odd mirror-parity is maximized \citep{Ben-David:2012eu,Finelli:2012ij,Planck-Collaboration:2013oq} has been assigned much higher levels of statistical significance. The various works, however, differ greatly in the methods used to analyse the data, and utilize different statistical estimators, Galactic masks and significance estimation methods.

This work aims to shed a clear light on these findings. We use the recent \textit{Planck} data release to perform a robust search for mirror-parity using both a pixel-based statistic \citep{de-Oliveira-Costa:1996qy,de-Oliveira-Costa:2004fy} and a statistic in harmonic space \citep{Ben-David:2012eu, Finelli:2012ij}. We address several issues involved, from the statistical methods used in the analyses and their possibly inherent biases, to the choice of the CMB maps and foreground masks that are investigated. We also analyse the scale dependence of the results and compare them to results on random simulations which are deliberately manipulated so that they contain different types and amplitudes of mirror-parity symmetry.

The structure of this paper is as follows. In Section~\ref{sec:DataAndSimulations}, we describe the data maps used for the analysis, the foreground masks we apply and the set of random simulations we use for evaluating the significance of the results. In Section~\ref{sec:Method}, we present the two statistics we use to search for mirror-parity in the CMB data as well as our methods for significance estimation. In Section~\ref{sec:TestsOnRandoms}, we describe three tests we perform on our statistics in order to compare their effectiveness in detecting mirror-parity and their sensitivity to the power spectrum amplitude and the shape of the Galactic masks. In Section~\ref{sec:Results}, we present the results of our analysis on the real data and estimate their statistical significance. We conclude in Section~\ref{sec:Conclusion}.

\section{Data and Simulations}
\label{sec:DataAndSimulations}

Our main focus in this work is the first data release of CMB sky maps from the \textit{Planck} experiment \citep{Planck-Collaboration:2013kl}. We shall use two of the four component separation maps available -- the SMICA and NILC maps. These two maps are the cleaner of the four, and are accompanied by smaller Galactic masks. The other two component separation maps, SEVEM and Commander--Ruler, both give similar results and were left out for brevity.
The third map we use has recently been released by \citet{Bobin:2014qe}. It was created using the LGMCA component separation method, and it combines the data from the 9-yr release of \textit{WMAP} as well as the first \textit{Planck} release (we use the map \citealt{Bobin:2014qe} refer to as WPR1). It is claimed to be a rather clean full sky map, that can be trusted, like the \textit{Planck} SMICA and NILC maps, with small Galactic masks, and also completely unmasked. We refer to this map as LGMCA.

Since we are interested in analysing the large scales of the CMB sky, we smooth the data maps and degrade them to a low \textsc{HEALPix} \citep{Gorski:2005fk} resolution of $\Nside=16$, corresponding roughly to a harmonic scale of $\lmax=48$. This also gives the added benefit of significantly reducing the computation time required for the analysis. Each map is first deconvolved in harmonic space with the beam and pixel window functions. It is then convolved with a Gaussian beam with FWHM of $\sim 11\degr$, equivalent to 3 pixels of an $\Nside=16$ map. We retain only the harmonic coefficients up to $\lmax=64$ and convert the smoothed coefficients back to an $\Nside=16$ map.

We use three Galactic masks in our analysis, of various sizes, all released by the \textit{Planck} team \citep{Planck-Collaboration:2013kl}. As the main mask we use the fairly large U73 mask, which is the union of the confidence masks for the individual component separation maps. 
This mask covers 27 per cent of the sky. In addition, we apply two smaller masks to the SMICA map -- its confidence mask, covering 11 per cent of the sky, named CS-SMICA89 \citep{Planck-Collaboration:2013kl} and the small inpainting mask defined by the \textit{Planck} team for the SMICA map, which we denote as SMICA-INP. This mask covers only 3 per cent of the sky.

\begin{figure}
\centering
\includegraphics[width=\halfW]{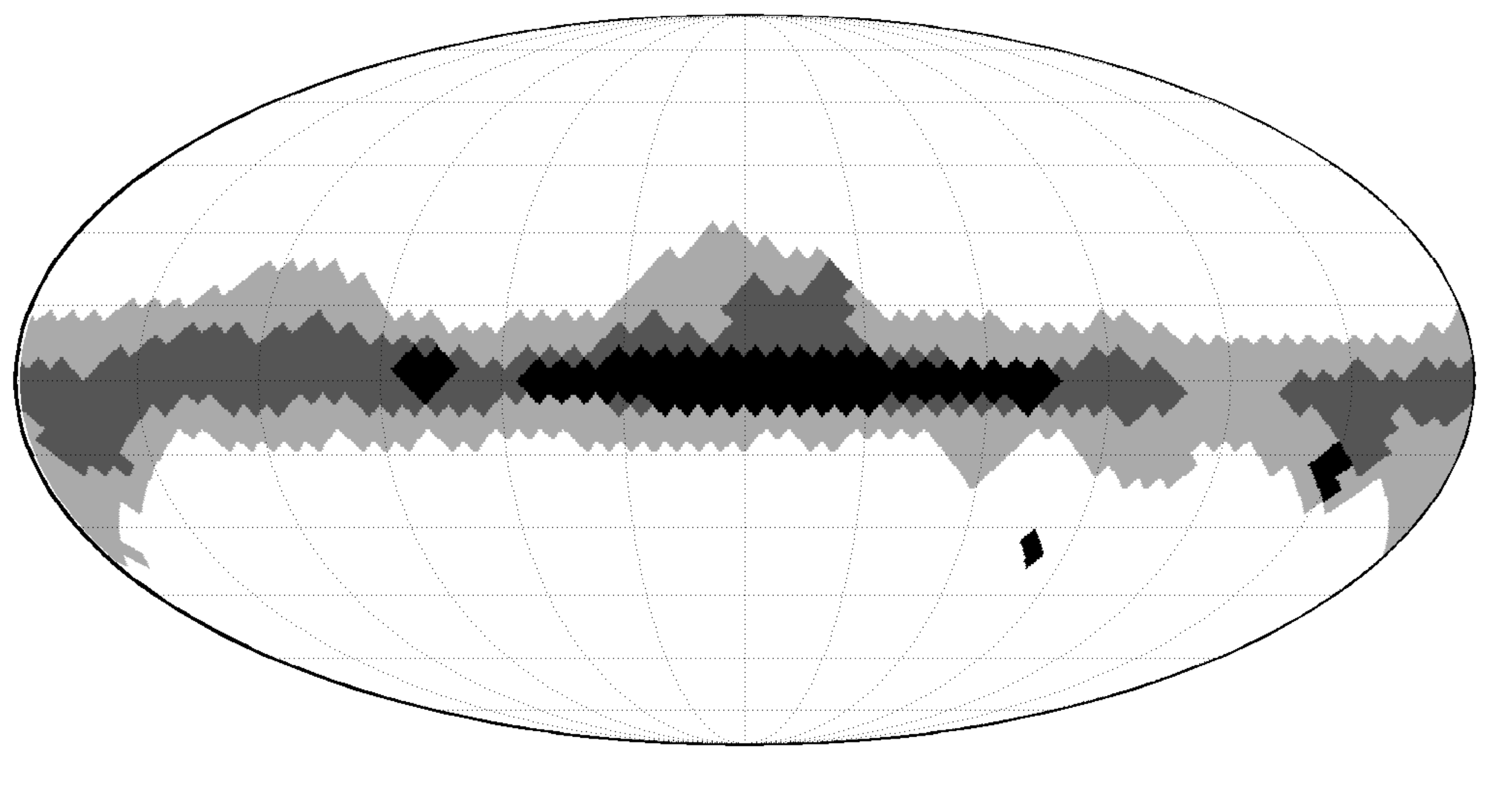}
\caption{The masks used in this work: U73 (\emph{light grey}), CS-SMICA89 (\emph{dark grey}) and SMICA-INP (\emph{black}), which have been smoothed, degraded to $\Nside=16$ resolution and thresholded as explained in the text.}
\label{fig:masks}
\end{figure}

The degradation of these masks to a low resolution is done in a similar manner to the process used for the data maps. The masks are first smoothed with the same Gaussian kernel as the data maps, and converted to $\Nside=16$ maps. In addition, they are thresholded to produce a binary mask. We have chosen the threshold for each mask so that the masked area will not be changed by the process of smoothing and degradation. The threshold values we use are $0.68, 0.57$ and $0.81$ for the U73, CS-SMICA89 and SMICA-INP masks, respectively. The resulting three low-resolution masks are shown in Fig.~\ref{fig:masks}.

In order to test the significance of our findings, we compare them to realizations of the statistically isotropic \LCDM. We use the \textit{Planck} maximum-likelihood power spectrum \citep{Planck-collaboration:2013xe} to generate $10^4$ random realizations with $\lmax=48$. The harmonic coefficients are smoothed with the same Gaussian kernel used to smooth the data maps and converted to $\Nside=16$ maps. No detector noise is added to these simulations (for the large scales we are studying, the dominant source of uncertainty is cosmic variance).
We note that when comparing the data with the simulations, we always treat the simulations in exactly the same way as the data, i.e., if the data is masked, we compare it to randoms which have been masked with the same mask. 

\section{Method}
\label{sec:Method}

In the various previous works testing the CMB sky for signs of mirror-parity, two different statistics were used. One, introduced by \citet{de-Oliveira-Costa:1996qy}, uses the data in pixel-space, and the other, introduced by \citet{Ben-David:2012eu}, uses the harmonic decomposition of the CMB sky. In this work we use both approaches. This allows us to compare between the two statistics and present a broader analysis.

\subsection{Pixel-Based Statistic}

Under mirror-parity, a direction $\bhat{r}$ in the sky transforms as
\be
\bhat{r}\to\bhat{r}_{\bhat{n}}=\bhat{r}-2\left(\bhat{r}\cdot\bhat{n}\right)\bhat{n},
\ee
where $\bhat{n}$ is the normal to the mirror plane. In order to check for mirror-parity in pixel-space, one can simply compare the sky hemispheres by using the statistic \citep{de-Oliveira-Costa:2004fy, Finelli:2012ij, Planck-Collaboration:2013oq}
\be \label{eq:PixelSpaceScore}
S_\rmn{p}^\pm(\bhat{n})=\overline{\left[\frac{T(\bhat{r})\pm T(\bhat{r}_{\bhat{n}})}{2}\right]^2},
\ee
where the over-bar denotes an average over all spatial directions $\bhat{r}$. With this statistic, a large degree of even (odd) mirror-parity is given by a low $S_\rmn{p}^-$ ($S_\rmn{p}^+$).

In the pixel-based approach, it appears that applying a Galactic mask is straightforward, as the masked pixels can simply be ignored when calculating the average. However, we will show below that this introduces an anisotropy bias to this statistic. As we shall demonstrate, an additional bias comes from the fact that this statistic is not normalized by the power spectrum amplitude.

\subsection{Harmonic Statistic}

The transformation of the spherical harmonics under mirror-parity through the $z$-axis yields
\be
Y_{\ell m}(\bhat{r})\to(-1)^{\ell+m}Y_{\ell m}(\bhat{r}).
\ee
We therefore use the natural statistic for mirror-parity in harmonic space \citep{Ben-David:2012eu}
\be \label{eq:HarmonicScore}
S_\rmn{h}(\bhat{n}) = \sum_{\ell=2}^\lmax\sum_{m=-\ell}^\ell (-1)^{\ell+m}\frac{\left| a_{\ell m}(\bhat{n}) \right|^2}{\widehat{C}_{\ell}} -(\lmax-1),
\ee
where $\bhat{n}$ is the $z$-axis used in the harmonic expansion, 
\be\label{eq:ActualPower}
\widehat{C}_\ell = \frac{1}{2\ell +1}\sum_m \left| a_{\ell m}\right|^2
\ee
is the \emph{actual} observed power spectrum and the subtraction of $\lmax-1$ simply ensures that the ensemble mean is zero. The free parameter, $\lmax$, that determines the smallest scale taken into account, allows us to easily test the scale dependence of the results. Since we are interested in testing the mirror-parity of large scales, in this work we mainly consider $\lmax=5,\dots,9$. Using the harmonic statistic, a large degree of even (odd) mirror-parity is given by a high (low) value for $S_\rmn{h}$.

Unlike in the pixel-based approach, here care must be taken when transforming the sky maps to harmonic space after applying a Galactic mask. Since the spherical harmonic functions are not orthogonal on the cut sky, one cannot simply perform a harmonic decomposition of the masked sky. The method we use for harmonic reconstruction is described in the appendix. 
The need for reconstruction of the harmonic coefficients when using a Galactic mask provides a limitation on the maximal value of $\lmax$. As the reconstruction errors for a given mask increase for smaller scales \citep{de-Oliveira-Costa:2006fp}, we conservatively limit the smallest scale we test using masked maps to $\ell=9$. When using unmasked data, we test also smaller scales, of up to $\lmax=48$, the rough limit allowed by our chosen resolution for the maps.

\subsection{Significance Estimation}

We use two methods in order to estimate the significance of the mirror-parity scores on the input data maps. 
In the first method, the `raw' score is used. We define $R$ to be the best score of each statistic, i.e., $R$ is the minimum over all directions of $S_\rmn{p}^-(\bhat{n})$ ($S_\rmn{p}^+(\bhat{n})$) for even (odd) parity using the pixel-based statistic, and the maximum (minimum) over all directions of $S_\rmn{h}(\bhat{n})$ for even (odd) parity using the harmonic statistic, 
\be
R=\left\{
\begin{array}{ll}
\min S_\rmn{p}^-(\bhat{n}) & \textrm{pixel-based even parity}, \\
\min S_\rmn{p}^+(\bhat{n}) & \textrm{pixel-based odd parity}, \\
\max S_\rmn{h}(\bhat{n}) & \textrm{harmonic even parity}, \\
\min S_\rmn{h}(\bhat{n}) & \textrm{harmonic odd parity}. 
\end{array} \right.
\ee
We then compare $R$ to its value on randoms simulations. Note that for the pixel-based statistic, this is the same estimator used by the \textit{Planck} team in their search for anomalous mirror-parity \citep{Planck-Collaboration:2013oq}.

The second method was first introduced in \citet{Ben-David:2012eu} \citep[see also][]{Rassat:2013th}. For each score map, we calculate its mean $\mu$ and standard deviation $\sigma$ over all directions. We then normalize the best score as
\be
\bar{R}=\frac{\left|R-\mu\right|}{\sigma},
\ee
and count the number of random simulations with a higher value of $\bar{R}$. This estimator tests how significant the best direction is relative to the rest of the score map, before comparing to simulations. It is particularly well suited for a scenario in which the anomalous mirror-parity is of cosmological origin, e.g.~induced by some large scale pre-inflationary effect.\footnote{An extension of the model introduced in \citet*{Fialkov:2010lr} \citep*[see also][]{Kovetz:2010fk}, in which a pre-inflationary particle in motion remains in the vicinity of the observable Universe after inflation, can be shown to induce traces of mirror-parity in the CMB sky.} In such a scenario, we expect to discover a \emph{single} pronounced direction of mirror-parity in the sky. The $\bar{R}$ estimator therefore complements the $R$ estimator, which only considers the best direction of each map.

\section{Tests on Random Simulations}
\label{sec:TestsOnRandoms}

Before using the pixel-based and harmonic statistics on the real CMB data, we test their performance on random simulations to compare their efficiency in detecting signs of mirror-parity and their sensitivity to the overall CMB power and to the use of large sky masks.

\subsection{Detectability of Mirror-Parity}

In order to test the ability of our statistics to detect anomalous mirror-parity in the CMB data, we use random realizations that have been modulated to contain different traces of mirror-parity. Since we are searching for any signs of mirror-parity, without limiting ourselves to a specific model, we modulate the random realizations using very simple phenomenological models. 

Before modulating each random map, we first choose a mirror-parity axis at random and use it as the $z$-axis to perform a harmonic decomposition of the map. We then rescale each of the harmonic coefficients $a_{\ell m}$ by the non-negative factors $f_{\ell m}$ given by
\be
f_{\ell m} = \left\{
\begin{array}{ll}
e_\ell & \ell+m \textrm{ is even},\\
o_\ell & \ell+m \textrm{ is odd}.
\end{array} \right.
\ee
The total power is kept constant by the requirement that these even and odd coefficients satisfy
\be\label{const_power}
(\ell+1)e_\ell^2 + \ell o_\ell^2=2\ell+1
\ee
for each $\ell$. We introduce a scale dependent amplitude $x_\ell \in [-1,1]$, with $x_\ell=1$ indicating maximal even parity and $x_\ell=-1$ maximal odd parity. To modulate with even (odd) mirror-parity, we set $o_\ell = 1-x_\ell$ for $x_\ell\ge0$ ($e_\ell=1+x_\ell$ for $x_\ell<0$), and use the power conservation equation \eqref{const_power} to determine $e_\ell$ ($o_\ell$).

We perform the modulation according to two different schemes:
\begin{enumerate}
\item Constant mirror-parity, for which the amplitude is constant for all scales, $x_\ell=x$.
\item Decaying mirror-parity, for which the amplitude is exponentially decaying. We set a pivot scale $\ell_\star=7$ and, for a given global parity amplitude $x$, set the parity level as $x_\ell=\rmn{sgn}(x)|x|^{\ell/\ell_\star}$.
\end{enumerate}
We vary the mirror-parity amplitude $x$ and examine simulations with even mirror-parity at amplitudes $x=0.1,0.2,\dots,0.8$. Apart from this mirror-parity modulation, these simulations are created in exactly the same manner as the clean \LCDM\ simulations discussed above. We draw $10^4$ realizations and modulate each of them with the eight amplitude values, in each of the two modulation schemes.

\begin{figure*}
\centering
\includegraphics[width=\fullW]{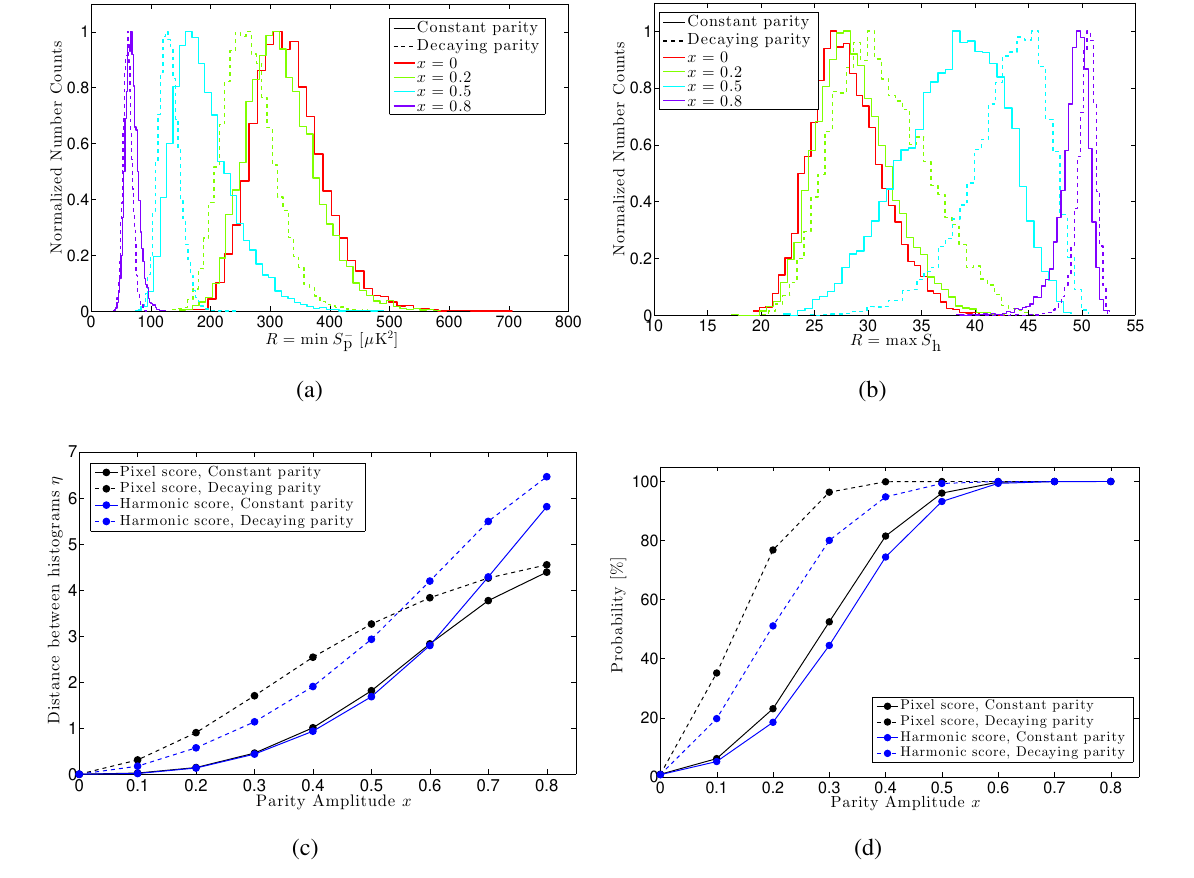}
\caption{Effectiveness of the different parity statistics, measured using the $R$ estimator on random realizations modulated with various amplitudes of even parity. All harmonic scores are calculated here with $\lmax=7$. (a)~Histograms of $R$ for the pixel-based statistic. (b)~Histograms of $R$ for the harmonic statistic. (c)~The normalized distance between modulated and unmodulated histograms, as defined in Eq.~\eqref{eq:HistDist}. (d)~The probability of detecting the parity modulation based on the distance between the modulation axis and the direction of the best parity score, as described in the text.}
\label{fig:HidingResults}
\end{figure*}

The results of testing the effectiveness of the parity estimators using the modulated random realizations are plotted in Fig.~\ref{fig:HidingResults}.
Panels~(a) and~(b) show histograms of the $R$ estimator for different modulation amplitudes, in both the constant and decaying schemes, for the pixel-based and harmonic statistics, respectively. We see that as the parity amplitude is increased, the histograms shift further away from the unmodulated ($x=0$) simulations.

This is summarized in panel~(c), where we plot for each modulation amplitude $x$ the normalized distance between its histogram and the unmodulated histogram. We define this distance as
\be\label{eq:HistDist}
\eta(x)=\frac{\left|\mu(x)-\mu(0)\right|}{\sqrt{\sigma^2(x)+\sigma^2(0)}},
\ee
where $\mu(x)$ and $\sigma(x)$ are the mean and standard deviation of the $R$ estimator, respectively. This distance can be thought of as a rough estimate of the signal-to-noise ratio for the detection of such mirror-parity. We see that for both the constant and decaying modulation schemes, both types of statistics, the pixel-based and harmonic, perform very similarly, the former being slightly more efficient. 

While so far we have tested the detectability of the mirror-parity modulation using the \emph{value} of the $R$ estimator, we can also use the \emph{direction} as our testing tool. For each modulated random realization, we calculate the distance between the parity axis used to modulate the map and the direction which yields the best parity score $R$. If this distance is smaller than $8\degr$ (i.e., within two $\Nside=16$ pixels), we consider this a `successful' detection of the modulated parity. We count the number of realizations yielding a detection, and plot in Fig.~\ref{fig:HidingResults}(d) the probability for detecting the modulation by this direction-based test. We see that this test also shows that both estimators are effective in detecting mirror-parity with a high enough amplitude. Again, we see that the pixel-based statistic performs better than its harmonic counterpart.

Finally, comparing between panels~(c) and~(d), we see that both the intensity-based and direction-based tests yield similar results. The decaying modulation scheme is in general more easily detected than the constant scheme.\footnote{Comparing between the two schemes is not straightforward, though, as the $\ell_\star=7$ pivot scale for the decaying scheme has been arbitrarily chosen.}

\subsection{Sensitivity to the Total Power}
\label{subsec:ConstrainedRandoms}

When searching for some trait or property in the data, it is desirable to devise a statistic that is as sensitive as possible to this property, but that is also as insensitive as possible to any other property. Otherwise, the statistical significance of any finding is `contaminated' by residual significance levels of other properties and oddities picked up by the statistic.

Following this reasoning, we wish to disentangle the question of the existence of anomalous mirror-parity in the CMB from that of the normality or abnormality of the total measured power on large scales. Since mirror-parity is a trait related to the relative distribution of the phases of harmonic coefficients, their total power is irrelevant. As a lack of power on large scales compared to the \LCDM\ expectation is evident (though not significant by itself, see \citealt{Bennett:2011lr,Planck-collaboration:2013xe}) in the \textit{Planck} maps, such entanglement could have an important effect on the search for mirror-parity. 

Upon examination of our two statistics, Eqs.~\eqref{eq:PixelSpaceScore} and \eqref{eq:HarmonicScore}, we can naturally expect the harmonic statistic to be insensitive to the total power of each scale, as it is specifically normalized by this factor, $\widehat{C}_\ell$ of Eq.~\eqref{eq:ActualPower}. The pixel-based statistic, however, may be sensitive to this property.

To test that, we generate a third set of $10^4$ simulations. Following \citet{Copi:2013qv}, we generate constrained realizations of the measured \LCDM\ power spectrum. For each realization, we draw random values $\widetilde{C}_\ell$ from Gaussian distributions centred around the measured values, with standard deviations corresponding to the measurement errors. This produces a power spectrum consistent with the one measured by the experiment, and allows us to produce realizations of \emph{our} Universe, as reported by observations, instead of realizations of the full \LCDM\ model. As stressed in \citet{Copi:2013qv}, for this aim cosmic variance is irrelevant, as is the fact that the $C_\ell$ are $\chi^2$-distributed in \LCDM. We use the \textit{Planck} maximum-likelihood power spectrum \citep{Planck-collaboration:2013xe}, available unbinned for the large scales ($\ell\leq48$) we require. 
As the \textit{Planck} measurement errors were not made available separate from the cosmic variance component, we use the (larger) measurement errors reported by \textit{WMAP} \citep{Larson:2011it} instead, but this choice has a negligible effect on our results. 
We then constrain a given set of $a_{\ell m}$ coefficients by calculating their power $\widehat{C}_\ell$ using Eq.~\eqref{eq:ActualPower} and rescaling them by the factors $\sqrt{\widetilde{C}_\ell/\widehat{C}_\ell}$. Each random set of $\widetilde{C}_\ell$ is used to create a single constrained realization. Apart from constraining their power, these simulations are created in exactly the same manner as our clean \LCDM\ simulations.

This set of constrained realizations can now be used instead of the normal set to compute the significance level of either the $R$ or $\bar{R}$ parity estimators. We wish to measure how much the significance level for each estimator is influenced by the change of ensembles, assuming the data indeed show a tendency for mirror-parity. To do so, we use random realizations which have been mildly modulated with even mirror-parity (amplitude $x=0.2$), either in the constant or decaying modulation schemes. We take the median\footnote{The median is more appropriate here than the mean, since the statistics are non-Gaussian and slightly skewed, as can be seen in Fig.~\ref{fig:HidingResults}(a)--(b).} of the score over all modulated realizations using each of the mirror-parity estimators, and calculate its significance level $\sigma$ compared to either the constrained or unconstrained random realizations. The differences between the two, $\Delta\sigma$, are summarized in Table~\ref{tab:ConstCompareWithHiding}.
\begin{table}
\centering
\caption{The difference $\Delta\sigma$ between the significance level of the median score of random realizations modulated with even parity of amplitude $x=0.2$, when compared to the constrained ensemble and to the unconstrained ensemble. The harmonic scores are calculated with $\lmax=7$.}

\begin{tabular}{llcc}
\hline
& & Pixel & Harmonic \\
\hline
\multirow{2}{*}{Constant} & $R$  & 0.703 & 0.025 \\
& $\bar{R}$  & 0.537 & 0.008 \\
\hline
\multirow{2}{*}{Decaying} & $R$  & 0.431 & 0.027 \\
& $\bar{R}$  & 0.616 & 0.016 \\
\hline
\end{tabular}
\label{tab:ConstCompareWithHiding}
\end{table}

We see that while the harmonic statistic is hardly affected by the difference in ensembles and the significance level changes by only $\sim0.02\sigma$, the pixel-based statistic is very sensitive to this change. When comparing to the `wrong' ensemble, this statistic can give a significance level that is off by $\sim0.5\sigma$. These results do not vary considerably between the two modulation schemes and the two significance estimators. They are also not strongly affected by the modulations amplitude $x$, as long as it is not very high.

Since the large scales that we are testing here suffer from a large cosmic variance, the power spectrum from which they were drawn is not well known. We have been using the best-fitting \LCDM\ power spectrum to draw random realizations out of the assumption that this model is valid for all cosmological scales. However, effects that could change the power spectrum on large scales cannot, at this time, be completely ruled out by the data. We therefore conclude that when searching for evidence of mirror-parity in the CMB, the harmonic statistic, which is insensitive to the total power (by construction) is preferred over the pixel-based statistic, that can cause a notable over- or under-estimation of the significance of the results. 

\subsection{Bias on Masked Sky}
\label{subsec:MaskingBias}

Our final test on the two mirror-parity statistics is for bias due to the shape of the Galactic masks. The introduction of a mask over the sky breaks the isotropy of the search for mirror-parity, and can bias the process of evaluating the significance of the results (\citealt*{Inoue:2008uq}; \citealt{Copi:2011fr}). Even in a completely isotropic setup, on a masked sky some directions can be more likely to get an extreme parity score than others. This concern is worsened by the fact that Galactic masks tend to be roughly band-shaped and highly parity-even in Galactic coordinates \citep{Inoue:2008uq,Copi:2011fr}.

\begin{figure*}
\centering
\includegraphics[width=\fullW]{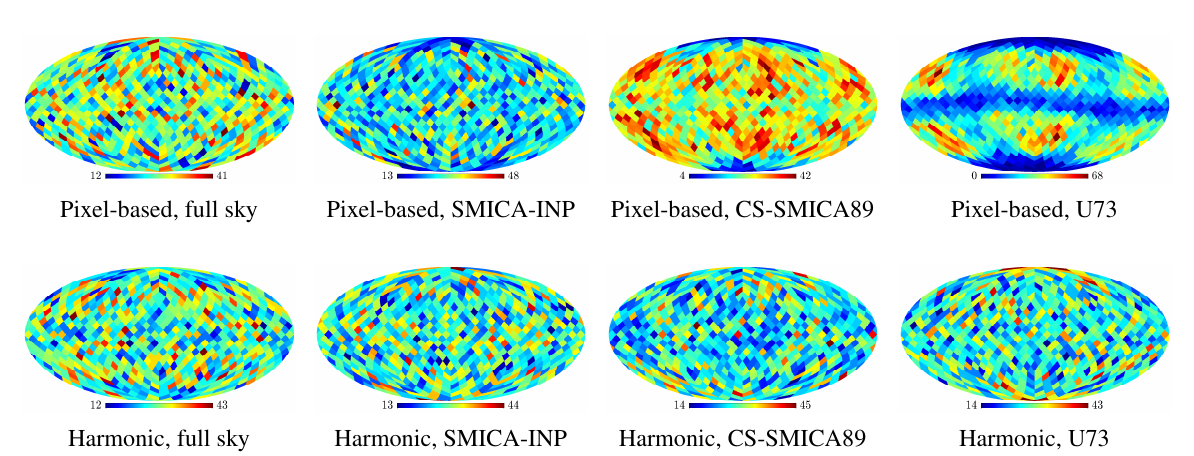}
\caption{Histograms of the direction of maximal even parity for $10^4$ randoms. The top row is for the pixel-based statistic, and the bottom row is for the harmonic statistic, with $\lmax=7$. The mask used on the random simulations is stated for each panel. Since the mirror-parity statistics are symmetric and obey $S(-\bhat{n})=S(\bhat{n})$, the histograms were calculated on half of the sky and duplicated to the other half, for visualization purposes.}
\label{fig:IsotropyHistograms}
\end{figure*}

We therefore calculate, for both the pixel-based and harmonic parity statistics, the direction exhibiting maximal even parity for each of our $10^4$ random simulations. We do this for unmasked simulations, as well as after masking the simulations with each of the three masks -- SMICA-INP, CS-SMICA89 and U73. In Fig.~\ref{fig:IsotropyHistograms}, we plot histograms of the results.

Since the random simulations are realizations of a completely isotropic statistical model, any deviation of the results from isotropy must be associated with the Galactic mask, and shows the sensitivity of the parity statistic to this mask. We can calculate how consistent each histogram of Fig.~\ref{fig:IsotropyHistograms} is with a uniform distribution on a sphere using, e.g., a simple $\chi^2$ test. 
We find that on the full sky, without masking, both the pixel-based and the harmonic statistics are isotropic, as expected. The pixel-based statistic, however, becomes less and less isotropic as the masked area is increased. With the small SMICA-INP mask its histogram is marginally isotropic, while for the larger CS-SMICA89 and U73 masks it is significantly anisotropic. The harmonic statistic, however, remains isotropic on the masked sky, even when a large mask such as U73 is used. This is due to our use of the maximum likelihood method to reconstruct the full sky harmonic coefficients before calculating the score. It is evident now that for the large scales we require here, the reconstruction method does not introduce a bias to the phases of the coefficients, leaving the harmonic statistic free of bias due to masking.

In summary, before moving on to the results on the real CMB data, we have shown that while both pixel-based and the harmonic statistics are equally efficient in detecting CMB mirror-parity, the pixel-based statistic suffers from two major drawbacks compared to the harmonic one: it is sensitive to the total power on large scales, an effect that should be investigated and evaluated separately from mirror-parity, and it suffers from an anisotropy bias when a Galactic mask is used. Both effects alter the significance of the results. We therefore conclude that the harmonic statistic, which does not suffer from these two effects, should be preferred when analysing the CMB in search for signs of anomalous mirror-parity. The ability to investigate the scale dependence of the results via the $\lmax$ parameter is another added value of the harmonic statistic, contributing yet another reason for its preference over its pixel-based counterpart.

\section{Results}
\label{sec:Results}

For all map and mask combinations and for both statistics, we identify a single direction of maximal even mirror-parity, $(l,b)\sim(260\degr,48\degr)$, and a corresponding one for odd mirror-parity, $(l,b)\sim(264\degr,-17\degr)$ in the CMB data.
We now turn to the analysis of the significance levels of these findings using the two estimators discussed above, $R$ and $\bar{R}$.

\subsection{Using the Pixel-Based Statistic}

\begin{figure}
\centering
\includegraphics[width=\halfW]{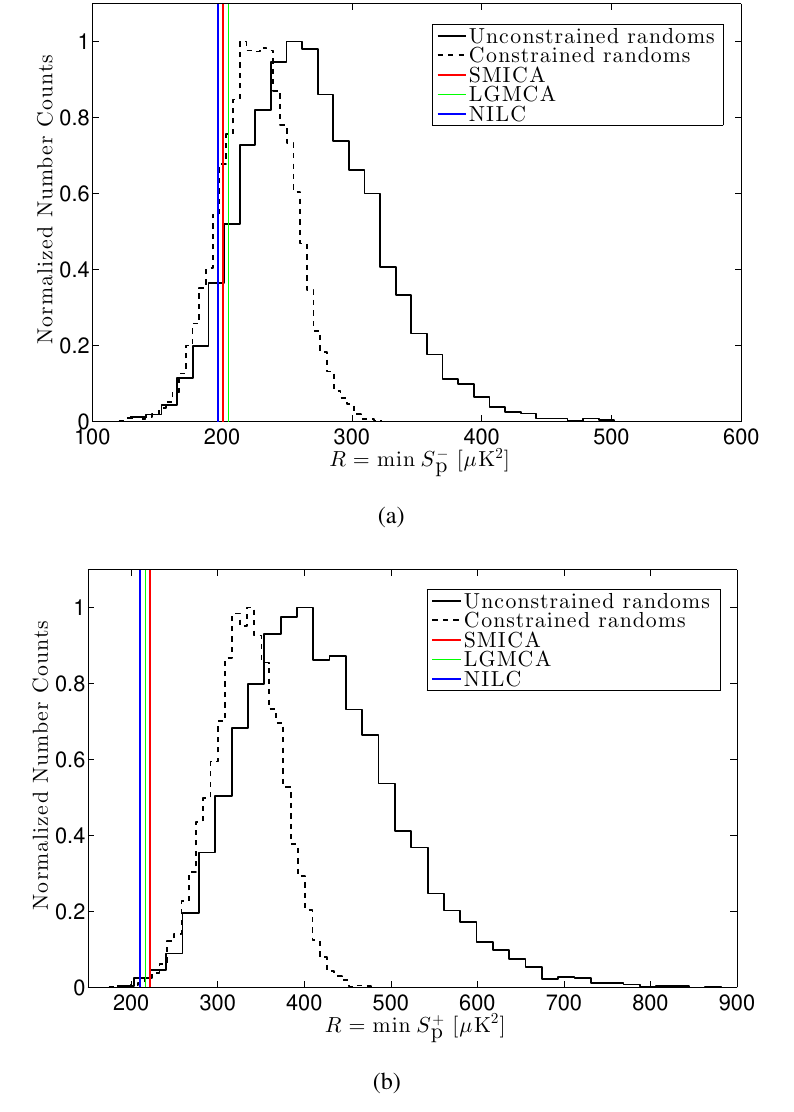}
\caption{The pixel-based results using the U73 mask and the $R$ estimator, for (a)~even and (b)~odd mirror-parity. The values for the various maps are the coloured vertical lines. Apart from the histograms of the results on our set of random simulations (\emph{solid}), we also plot the histograms of the results on the set of constrained realizations of Section~\ref{subsec:ConstrainedRandoms} (\emph{dashed}). All randoms are also masked with the U73 mask.}
\label{fig:PixelSpaceResults}
\end{figure}

In Fig.~\ref{fig:PixelSpaceResults}, we plot the results of the pixel-based statistic for our set of maps masked with the U73 mask using the $R$ estimator, along with a histogram of the results on our set of random simulations, masked with the same mask.
This figure closely resembles the corresponding fig.~36 of the first preprint version of \citet{Planck-Collaboration:2013oq}, in which the same analysis was performed using the same large mask. Considering panel~(a), it is immediately clear that none of the maps show any significant evidence of even mirror-parity. Indeed, regardless of which map and mask are considered, using either the $R$ or $\bar{R}$ estimators, no significance level reaches above $3\sigma$ (and most combinations do not even reach $2\sigma$). 

When inspecting panel~(b), the results for odd mirror-parity appear to be more significant. We summarize the odd mirror-parity significance levels for all the maps and masks in Table~\ref{tab:PixResOdd}.
\begin{table}
\centering
\caption{Pixel-based significance levels (in $\sigma$) for odd mirror-parity. The same mask is used for both the data and random simulations.} 
\begin{tabular}{llcc}
\hline
Map & Mask & $R$ &  $\bar{R}$  \\
\hline
\multirow{4}{*}{SMICA} & U73 & 3.01  & 2.42 \\
 & CS-SMICA89 & 2.46  & 2.31  \\
 & SMICA-INP & 2.54 & 2.81  \\
& --- & 1.92  & 1.90  \\
\hline
\multirow{2}{*}{LGMCA}  & U73 & 3.16  & 2.28  \\
 & --- & 2.33  & 2.56  \\
\hline
\multirow{2}{*}{NILC}  & U73 & 3.29  & 2.54  \\
& --- & 2.64  & 2.89  \\
\hline
\end{tabular}
\label{tab:PixResOdd}
\end{table}
We see that as Fig.~\ref{fig:PixelSpaceResults}(b) suggests, the significance levels for the maps masked with the U73 mask are somewhat high when measured using the $R$ estimator, crossing the $3\sigma$ level. However, when the same results are evaluated using the $\bar{R}$ estimator, the significance level already drops to $\sim2.5\sigma$, for all maps. Furthermore, when a smaller mask is used, or when the maps are tested without a Galactic mask, the significance levels never reach $3\sigma$ on any of the maps. This is in general agreement with the results reported in the updated preprint version of \citet{Planck-Collaboration:2013oq}, in which the $R$ estimator is used on the full sky and not on maps masked with the U73 mask as in the first version \citep[see updated version of][fig.~40]{Planck-Collaboration:2013oq}. As was demonstrated above (see Section~\ref{subsec:MaskingBias}), results of the pixel-based statistic on the masked sky are inherently biased, especially with a mask as large as U73. The change in analysis method between versions of \citet{Planck-Collaboration:2013oq} suggests that this has also been realized by the Planck Collaboration. Here, we have demonstrated this effect explicitly.

Finally, we refer the reader to the second histogram plotted in both panels of Fig.~\ref{fig:PixelSpaceResults}, showing the results on the set of constrained realizations of Section~\ref{subsec:ConstrainedRandoms}. This is done as a final demonstration of the effect of the total power of the pixel-based statistic. We can easily see how the same results of the CMB data maps are assigned a lower significance level using the set of constrained realizations, as the corresponding histograms are shifted towards smaller values. 
This sensitivity hampers our ability to evaluate the mirror-parity level separately from other effects by using the pixel-based statistic. As we have demonstrated above, the harmonic statistic is more robust and does not suffer from this disadvantage.

\subsection{Using the Harmonic Statistic}

Much like the pixel-based statistic, the harmonic statistic attributes very low significance levels to the even mirror-parity results on all combinations of maps and masks, using both the $R$ and $\bar{R}$ estimators.

In Table~\ref{tab:HarmRes}, we present the significance levels of the odd mirror-parity search results, using the harmonic statistic.
\begin{table}
\centering
\caption{Harmonic significance levels (in $\sigma$) for odd parity, with $\lmax=7$. Random simulations are always masked with the same mask as the data.} 
\begin{tabular}{llcc}
\hline
Map & Mask & $R$ & $\bar{R}$ \\
\hline
\multirow{4}{*}{SMICA} & U73 & 1.31  & 0.96  \\
 & CS-SMICA89 & 1.91  & 1.52  \\
 & SMICA-INP & 1.89  & 1.50  \\
& --- & 1.88  & 1.50  \\
\hline
\multirow{2}{*}{LGMCA}  & U73 & 1.47  & 1.02  \\
& --- & 2.22 & 1.88  \\
\hline
\multirow{2}{*}{NILC}  & U73 & 2.53 & 2.40  \\
& --- & 2.57  & 2.13  \\
\hline
\end{tabular}
\label{tab:HarmRes}
\end{table}
We include the results of both estimators for all the maps and masks in the same format as in Table~\ref{tab:PixResOdd}. We can immediately see that no significance level in Table~\ref{tab:HarmRes} is higher than $2.57\sigma$. Furthermore, while the LGMCA and NILC maps reach significance levels higher than $2\sigma$, the SMICA map yields very low levels, regardless of which mask is used.

\begin{figure}
\centering
\includegraphics[width=\halfW]{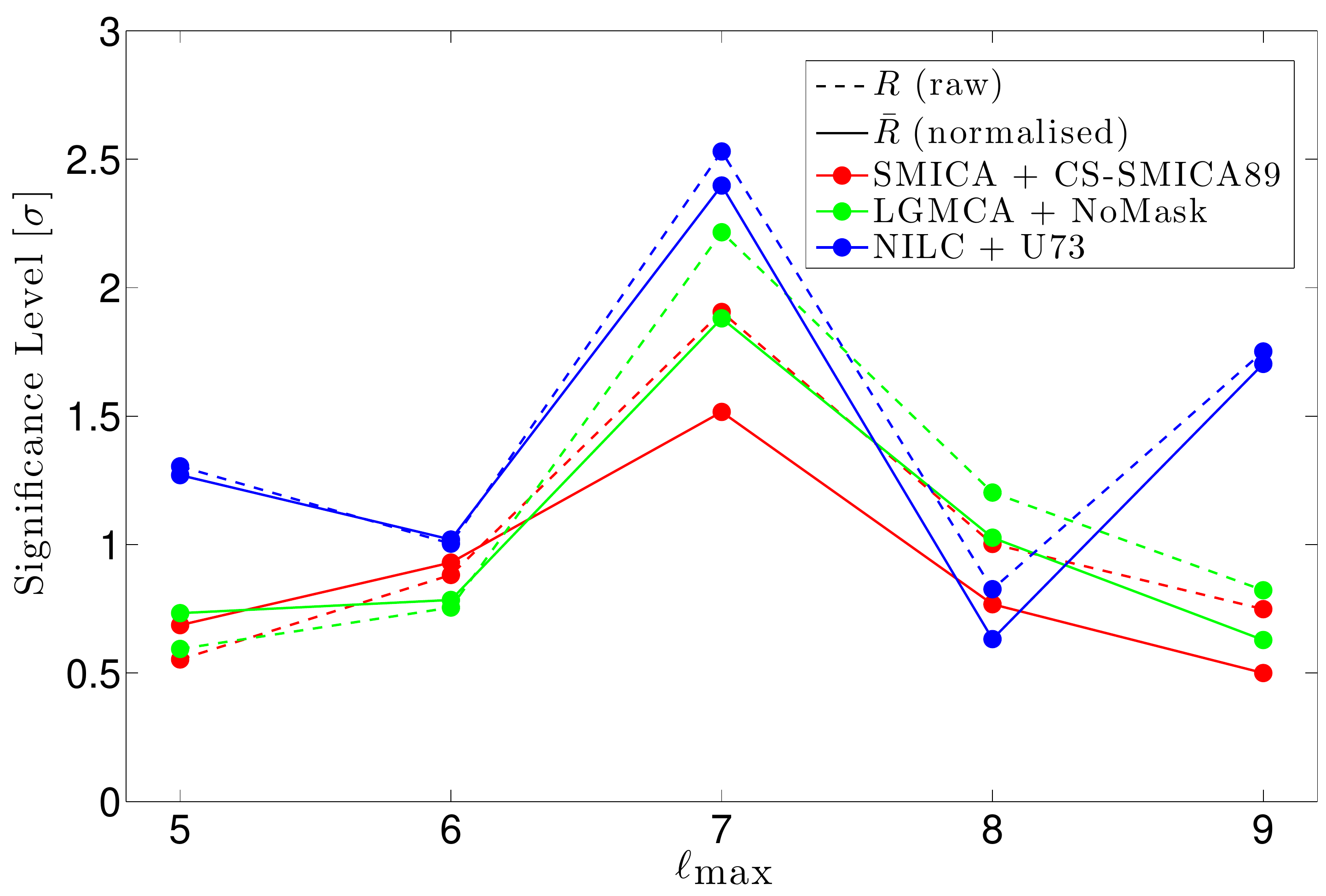}
\caption{The significance levels of odd parity for the harmonic statistic using both the $R$ (\emph{dashed}) and $\bar{R}$ (\emph{solid}) estimators, as a function of scale $\lmax$. As examples, we plot the results for the SMICA map masked with CS-SMICA89 (\emph{red}), the unmasked LGMCA map (\emph{green}) and the NILC map masked with U73 (\emph{blue}).}
\label{fig:HarmonicSpaceResults}
\end{figure}

The results shown in Table~\ref{tab:HarmRes} have been calculated while setting $\lmax=7$ in Eq.~\eqref{eq:HarmonicScore}, for both the data maps and random simulations. However, the analysis in harmonic space further allows us to check the scale dependence of the results by changing the $\lmax$ parameter. We therefore plot in Fig.~\ref{fig:HarmonicSpaceResults} the significance levels of the results as a function of $\lmax$ for some examples of map and mask combinations.
We can see that even the mildly significant map and mask combinations presented in Table~\ref{tab:HarmRes} become completely insignificant when $\lmax$ of the harmonic statistic is changed. The sensitivity of the results to the scale considered appears to be very strong. Furthermore, we see no apparent trend in any of the plotted lines. A cosmologically originated anomalous parity due to a pre-inflationary effect, for example, is expected to produce a decaying significance level when plotted against decreasing scales.

As previously mentioned, when choosing $\lmax$, we are limited by the harmonic reconstruction and therefore do not test scales smaller than $\ell=9$ on masked maps. On full sky maps, however, the limiting scale is determined by the resolution of the maps. When testing the unmasked maps with every $\lmax$ up to $\lmax=48$, we find no rise in the significance of odd parity and no apparent trend. Moreover, we note that the direction of maximal odd parity is no longer stable and changes considerably when smaller scales are tested. This further supports the notion that the level of odd mirror-parity is consistent with that of a random fluctuation.

\section{Discussion and Conclusion}
\label{sec:Conclusion}

Over the past decade, the pursuit of large-scale anomalies in CMB data has generated numerous claims of deviations from the expected behaviour according to the concordance \LCDM\ model \citep[see][and references within]{Bennett:2011lr,Planck-Collaboration:2013oq}, spurring a tumultuous discussion of the statistical methods used in the analyses \citep{Bennett:2011lr} and especially the approaches towards significance estimation. Some of the disagreements are grounded in principles, such as the age-long Bayesian vs.\ frequentist debate, and are likely to persist. Certain claims might be considered as questions of taste. For instance, some advocate that the statistical significance of each new reported result be normalized according to the total number of tests conducted on CMB data hitherto. This may be viewed as an attempt to compensate for the `look elsewhere' effect, in the sense of looking for many anomalies on a given data set (as opposed to the effect in the sense of the estimation of the significance of a particular one), although it is not clear how this normalization is to be tracked and whether it should carry over to each new data set. Regardless of personal convictions, it is important that results of great potential importance such as those cited above undergo careful scrutiny and robust examination if they are to advance our theoretical understanding of cosmology in a meaningful way.

In this work we have scrutinized previous claims of significant levels of mirror-parity in CMB maps from both the \textit{WMAP} and \textit{Planck} experiments. Indeed, using a pixel-based statistic and the large U73 Galactic mask, one can estimate the significance of the odd-mirror-parity direction to be as high as $3\sigma$ (0.13, 0.08 and 0.05 per cent for the SMICA, LGMCA and NILC maps, respectively). However, we have shown that these results are biased due to the large mask. With a smaller mask the significance level is lower, ranging from 0.25 to 1.04 per cent for the SMICA map. In addition, we have shown that these result are sensitive to assumptions regarding the total power on large scales, which is weakly constrained by the data. 

We have also tested the data for mirror-parity using a harmonic statistic. We have shown this statistic to be far more robust than its pixel-based counterpart -- stable against the use of a Galactic mask and with regard to the power spectrum amplitude. The statistical significance of the harmonic results reaches  a level of $\sim2.5\sigma$ (0.51 per cent) at its highest. This level, however, is sensitive to the choice of component separation method applied to the CMB data. In addition, we have shown that the significance level is highly sensitive to scale.

Furthermore, with the high quality of the \textit{Planck} maps, and especially the LGMCA map, we have also allowed ourselves to test the sky maps completely unmasked. This way, as long as the maps are clean enough of Galactic foregrounds, both statistics are expected to perform the best, free of anisotropy bias in the case of the pixel-based statistic (though still suffering from sensitivity to the total power) and of the \LCDM\ assumptions used for harmonic reconstruction in the case of the harmonic statistic. The results for the unmasked maps, however, are not statistically significant, with significance levels reaching only as high as 0.19 per cent for the pixel-based statistic and 0.51 per cent for the harmonic statistic. The pixel-based results are in general agreement with those of \citet{Planck-Collaboration:2013oq}.

In light of these findings, we conclude that while there is some tendency for odd parity in the CMB data, which peaks at the scale of $\ell=7$, when embracing a broader perspective and examining the complete set of data maps and Galactic masks and the properties of the statistical estimators, it appears that the evidence for anomalous mirror-parity is rather weak. Our conclusion is that it poses no real challenge to the concordance model, and should therefore not be considered a \LCDM\ anomaly.

\section*{Acknowledgements}

ABD thanks Pavel Naselsky for useful discussions. We also thank the anonymous referee for helpful comments. We acknowledge the use of the Legacy Archive for Microwave Background Data Analysis\footnote{http://lambda.gsfc.nasa.gov/} (LAMBDA) and the \textit{Planck} Legacy Archive\footnote{http://pla.esac.esa.int/pla/aio/planckProducts.html} (PLA). ABD was supported by the Danish National Research Foundation. EDK was supported by the National Science Foundation under Grant Number PHY-1316033.
 
{\footnotesize
	\bibliographystyle{mn2e_Fixed}
	\bibliography{MirrorParity}
}

\appendix
 
\section{Reconstruction of Large Scales}
\label{app:Reconstruction}

Several approaches can be found in the literature to the question of estimating the full-sky harmonic coefficients from a masked sky. These include direct inversion of the harmonic coupling matrix \citep*{Efstathiou:2010db,Bielewicz:2013nr}, Gaussian inpainting methods based on constrained realizations (\citealt{Inoue:2008uq}; \citealt*{Kim:2012jk}; \citealt{Copi:2013xy}), filtering methods with or without preconditioning \citep*{Smith:2007fk,Elsner:2013qy} and sparsity-based techniques \citep*{Starck:2013ty}.

Since we are only interested in the largest scales in this work, we have chosen to use the maximum likelihood method \citep{de-Oliveira-Costa:2006fp, Efstathiou:2010db, Aurich:2011nx, Ben-David:2012eu} to reconstruct the harmonic coefficients. The behaviour of this method has been discussed extensively in the literature (\citealt{de-Oliveira-Costa:2006fp, Efstathiou:2010db}; \citealt*{Feeney:2011kx}; \citealt{Copi:2011fr}). Following \citet{Feeney:2011kx} we use a Gaussian smoothing kernel instead of a top-hat, as this leads to superior performance on component separated maps.

In the maximum likelihood reconstruction method, the CMB data on the masked sky are represented as
\be
\bm{x} = \B{Y}\bm{a}+\bm{n},
\ee
where $\bm{x}$ is a vector of the temperature in the valid pixels, $\B{Y}$ is a matrix of the spherical harmonics\footnote{Without loss of generality, we use in this work the \emph{real} form of both the spherical harmonics and $a_{\ell m}$ coefficients. See \citet{de-Oliveira-Costa:2004fy}.} evaluated on each valid direction ($\B{Y}_{ij}=Y_{\ell_j m_j}(\bhat{r}_i)$), $\bm{a}$ is a vector of the harmonic coefficients we wish to reconstruct (i.e., containing only the scales $2\le\ell\le\ell_\rmn{rec}$) and $\bm{n}$ is a vector representing everything else that contributes to $\bm{x}$ -- including the CMB data of smaller scales ($\ell>\ell_\rmn{rec}$) and the detector noise -- and acts as noise for the reconstruction process. Assuming Gaussianity, an estimator $\bhat{a}$ of $\bm{a}$, which is unbiased ($\langle\bhat{a}\rangle=\bm{a}$) and has minimal variance, can be calculated as $\bhat{a}=\B{W}\bm{x}$, where the reconstruction matrix is
\be
\B{W} = \left(\B{Y}^\rmn{T}\B{C}^{-1}\B{Y}\right)^{-1}\B{Y}^\rmn{T}\B{C}^{-1}
\ee
and $\B{C}=\langle\bm{nn}^\rmn{T}\rangle$ is the noise covariance matrix.

Under the \LCDM\ assumption of statistical isotropy for the smaller scales of $\ell>\ell_\rmn{rec}$, we can estimate $\B{C}$ as
\be
\B{C}_{ij}=\sum_{\ell=\ell_\rmn{rec}+1}^L \frac{2\ell+1}{4\pi}P_\ell(\bhat{r}_i\cdot\bhat{r}_j)\,w_\ell^2b_\ell^2C_\ell + N\delta_{ij}, 
\ee
where $P_\ell$ are the Legendre polynomials and $C_{\ell}$ is the best-fitting \LCDM\ power spectrum. $b_\ell$ is the Gaussian smoothing kernel, allowing us to cut the summation at $L=48$, and $w_\ell$ is the pixel window function. Diagonal regularisation noise of $N=2~\mu\rmn{K}^2$ is added to keep the covariance matrix from becoming singular and has little effect on modes that contribute to the inversion. We choose $\ell_\rmn{rec}=9$ when reconstructing the harmonic coefficients, since this is the smallest scale we use to test for mirror-parity in this work.

\bsp

\label{lastpage}

\end{document}